\documentclass[twoside]{article}
\usepackage{fleqn,espcrc2}

\usepackage{graphicx}
\newcommand{\pom}{I\!\! P}
\newcommand\la{\langle}
 \newcommand\ra{\rangle}
 \newcommand\beq{\begin{equation}}
 \newcommand\noi{\noindent}
 \newcommand\eeq{\end{equation}}
 \newcommand\beqn{\begin{eqnarray}}   
 \newcommand\eeqn{\end{eqnarray}}

\newcommand{\AmS}{{\protect\the\textfont2
  A\kern-.1667em\lower.5ex\hbox{M}\kern-.125emS}}

\title{From hard to soft diffraction and return}

\author{B.Z.~Kopeliovich\address[MCSD]{Max-Planck-Institut f\"ur Kernphysik,
Postfach 103980, 69029 Heidelberg, Germany\\
Joint Institute for Nuclear Research, Dubna, Russia}}

\begin{document}

\begin{abstract}
 The long standing mystery of smallness of diffractive dissociation of
hadrons to large effective masses (the Pomeron-proton cross section is only
$2\,mb$) witnesses that the gluonic clouds of valence quarks are so small
($r_0=0.3\,fm$) that soft interaction hardly resolves those gluons
(diffraction is $\propto r_0^4$). A color-dipole light-cone (LC) approach
is developed which incorporates a strong nonperturbative interaction of the
LC gluons. The energy dependent part of the total hadronic cross section is
calculated in a parameter-free way employing the nonperturbative LC wave
functions of the quark-gluon Fock states.  It rises with energy as
$s^\Delta$, and we predict $\Delta=0.17\pm 0.01$, as well as the
normalization. However, the energy independent part of the cross section
related to inelastic collisions with no gluon radiated (gluons are not
resolved)  cannot be calculated reliably and we treat it as an adjustable
parameter which is fixed fitting just one experimental point for total
cross section. Then the energy dependence of the total cross section (the
Pomeron part) and the elastic slope are fully predicted, as well as the
effective Pomeron trajectory in impact parameter space, in a good agreement
with data. These results naturally explain the $x$-dependence of the
diffractive DIS observed at HERA. Although diffraction is expected to be
dominated by soft interactions the observed effective $\Delta$ is about
twice as large as one (0.08) known for total cross sections.  Diffractive
excitations of large effective mass correspond to diffractive gluon
radiation and should be associated with our calculated $\Delta$.

\vspace{1pc}
\end{abstract}

\maketitle

\section{Diffraction: evidence for small gluonic spots in
the proton}

The cross section of single diffraction in hadronic
collisions ($pp$ unless specified) can be
expressed in terms of the so called Pomeron-proton total
cross section \cite{mueller,kaidalov},
\beq
\frac{d\sigma(pp\to pX)}
{dt\,dM_X^2} \propto
\sigma^{\pom p}_{tot}\ .
\label{1}
\eeq
 What should one expect for $\sigma^{\pom p}_{tot}$? If to think about the
Pomeron as a gluonic system its cross section contains the Cazimir factor
$9/4$ compared to the meson-nucleon cross section. Thus, one may expect
$\sigma^{\pom p}_{tot}\sim 40-50\,mb$. Such a naive estimate is, however,
in a strict contradiction with data which show that the Pomeron interacts
with a tiny cross section about $2\,mb$ at high c.m. energy $M_X$
\cite{kaidalov}. This is the well know puzzling smallness of the
triple-Pomeron coupling which was under intensive discussion back in 70s
and is still waiting for a dynamical explanation. 

Although the unitarity corrections suppress diffraction 
they cannot bridge such a large gap between the expectation and data. 
Indeed, nothing surprising comes out from the analysis \cite{kklp} of the
same data for the triple-Regge coupling $\pom\pom R$ much larger that
$\pom\pom\pom$. The corresponding Reggeon-proton cross section
$\sigma^{Rp}_{tot}\approx 20\,mb$ agrees with
the expectation, and the absorptive corrections do not disturb it much.

The solution for the puzzling smallness of large-mass diffraction was
suggested in \cite{kst2}. Due to color screening (color transparency)  the
Pomeron-proton cross section depends quadratically on the size of the
gluonic system.  Indeed, glueballs are expected to be bound tighter than
quarkonia \cite{nsvz}, therefore to have smaller radiuses. Even the simple
relation \cite{cnn} between the color triplet (octet)  string tension and
and the slope of Reggeon (Pomeron) trajectory, $\kappa =
1/(2\pi\alpha_{R(\pom)}^\prime)$, leads to the octet string tension which
is at least four times larger than one for a triplet string (even more
since the effective $\alpha_{\pom}^\prime$ is substantially increased by
the unitarity corrections). 

Thus, we should conclude that the strong nonperturbative interaction of
gluons is responsible for smallness of large mass diffraction.

It worth reminding that the Pomeron is not a particle, but a shadow of
inelastic processes, and the concept of Pomeron-proton cross section is ill
defined (except if the Pomeron is a true Regge pole).  We use this concept
here in order to impose a scale which shows that the diffractive cross
section is abnormally small.  In what follows we employ the light-cone (LC) 
QCD formalism which relates the observed suppression of diffraction to the
smallness of gluonic spots inside the proton.  Indeed, the observed
dependence of the cross section Eq.~(\ref{1}), $1/M_X^2$, can only be due
to diffractive radiation of gluons since they are vector particles.  The
interaction with the target should be sufficiently hard to resolve the
gluonic fluctuation in a valence quark.  The smaller is the gluon cloud of
a valence quark, the more difficult is to excite it.  If the LC wave
function of the proton is dominated by configurations in which gluons are
located around the valence quarks at small transverse distances $\sim r_0$,
then the amplitude of diffraction is proportional to $r_0^2$ and the cross
section of diffractive gluon radiation is suppressed as $r_0^4$. Therefore,
high-mass diffraction is a very sensitive tool to study the gluonic
structure of the proton.  The related formalism based on the LC Green
function approach is presented below. 

\section{From hard to soft diffraction}

The first correct pQCD calculation for large-mass diffraction cross section
in DIS was performed in \cite{bartels} in the LC dipole representation. The
diffractive cross section of gluon radiation $q\,N \to q\,G\,N$ is given by
Eqs.~(64) of \cite{kst2} (see simple derivation in Appendix A.2 of
\cite{kst2}),
 \beqn
&& \frac{M^2\,d\sigma(qN\to qGN)}{dM^2\,dq_T^2}
\biggr|_{q_T=0} \nonumber\\
&=& \frac{81}{1024\pi}\,
\int d^2\rho\,\biggl|\Psi_{qG}(\alpha,\vec\rho)\,
\sigma_{\bar qq}(\rho,\bar s)\biggr|^2_{\alpha\ll1}\ ,
\label{2}
 \eeqn
 where $q_T$ is the transverse momentum of the radiated gluon, $M$ is the
quark-gluon effective mass, $\alpha$ is the fraction of the initial LC
momentum of the quark taken away by the gluon;  $\sigma_{\bar qq}(\rho,s)$
is the cross section of interaction with a nucleon of a $\bar qq$ dipole
with separation $\rho$ and energy $s$.  The perturbative quark-gluon
(photon) LC wave function is derived in \cite{hir} (see also in
\cite{kst1}). Using it one ends up with diffraction which grossly
overestimates data for $pp\to pX$, as one might has expected in view of
discussion in previous section. 

Of course one should not apply pQCD
methods to the soft reaction, at small $q_T$ the final quark and gluon
cannot be treated as free particles since they should experience a strong
nonperturbative interaction comoving with a large transverse separation
along the light cone. Propagation of such a $qG$ pair from initial
longitudinal coordinate $z_1$ and separation $\vec\rho_1$ to final $z_2$, 
$\vec\rho_2$ is described by the Schr\"odinger equation \cite{kst2},
 \beqn
&& i\frac{d}{dz_2}\,G_{qG}(z_1,\vec\rho_1;z_2,\vec\rho_2)
=
\Biggl[\frac{m_q^2 - \Delta_{\rho}}{2\,p\,\alpha\,(1-\alpha)}
\nonumber\\ &+& 
V_{\bar qq}(z_2,\vec \rho,\alpha)\Biggr]
G_{qG}(z_1,\vec\rho_1;z_2,\vec\rho_2)\ .
\label{3}
 \eeqn
 The first term in square brackets is the kinetic one, 
the LC potential is real for propagation in vacuum and describes the
quark-gluon interaction. We choose it in the
oscillator form in order to solve (\ref{3}) analytically,
 \beq
{\rm Re}\,V_{qG}(z_2,\vec\rho,\alpha) =
\frac{b^4(\alpha)\,\vec\rho\;^2}
{2\,p\,\alpha(1-\alpha)}\ ,
\label{3.10}
 \eeq
 where $b^2(\alpha)=b_0^2+4\,b^2_1\,\alpha\,(1-\alpha)$.
Since we are interested in $\alpha \ll 1$, $b(\alpha)=b_0$.

As far as the Green function is known one can calculate the nonperturbative
LC wave function for a quark-gluon Fock state,
 \beqn
&&\Psi_{qG}(\vec\rho,\alpha)=
\frac{i\,\sqrt{\alpha_{s}/3}}
{2\pi\,p\,\alpha(1-\alpha)}\nonumber\\
&\times&
\int\limits_{-\infty}^{z_2}dz_1\,
\Bigl(\bar\chi\;\widehat\Gamma\chi\Bigr)\,
G_{qG}(z_1,\vec\rho_1;z_2,\vec\rho)
\Bigr|_{\rho_1=0}
\label{3.9}
 \eeqn
 where $\chi,\ \bar\chi$ are the initial or final quark spinors, and the
vertex function reads,
 \beqn
\widehat\Gamma &=& i\,m_q\alpha^2\,
\vec {e^*}\cdot (\vec n\times\vec\sigma)\,
 + \alpha\,\vec {e^*}\cdot (\vec\sigma\times\vec\nabla)
\nonumber\\ &-& 
i(2-\alpha)\,\vec {e^*}\cdot \vec\nabla\ ,
\label{3.6a}
 \eeqn

which has a
rather simple form,
\beq
\Psi_{qG}(\vec r,\alpha\ll 1)=
-\,\frac{2\,i}{\pi}\,\sqrt{\alpha_s\over3}\,\,
\frac{\vec e\,^*\cdot\vec r}{r^2}\,
e^{-r^2b_0^2/2}\ ,
\label{4}
 \eeq
 where $\vec e$ is the polarization vector of the gluon.
This wave function recovers the perturbative one \cite{kst1} in the limit
of $b_0\to 0$.

Comparing the cross section of diffractive gluon radiation with data one
arrives at a rather large value of $b_0=0.65\,GeV$ corresponding to
a short mean quark-gluon separation $r_0 = 1/b_0 = 0.3\,fm$.
This is one of the central results of \cite{kst2}.

\section{Small gluonic spots and the total 
\boldmath$pp$ cross section}

Using the same approach one can calculate the cross section of
gluon bremsstrahlung (nondiffractive) by a $\bar qq$ meson
\cite{kst2,k3p1,k3p2},
 \beqn
&&\sigma^{hN}_1 =
\int\limits_0^1 d\alpha_q \int d^2R\,\,
\Bigl|\Psi^h_{\bar qq}
(R,\alpha_q)\Bigr|^2
\nonumber\\ & \times & \,
{9\over4}\int\limits_{\alpha\ll 1}
\frac{d\alpha}{\alpha}
\int d^2r
\biggl\{\Bigl|\Psi_{\bar qG}(\vec R
+\vec r,\alpha)\Bigr|^2
\nonumber\\ & \times & \,
\sigma_{\bar qq}^N(\vec R +\vec r)
+ \Bigl|\Psi_{qG}(\vec r,\alpha)\Bigr|^2
\sigma_{\bar qq}^N(r) 
\nonumber\\ &-&
{\rm Re}\,\Psi_{qG}^*(\vec r,\alpha)\,
\Psi_{\bar qG}(\vec R +\vec r,\alpha)\,
\nonumber\\ & \times & \,
\Bigl[\sigma_{\bar qq}^N(\vec R +\vec r)+\sigma_{\bar qq}^N(r)-
\sigma_{\bar qq}^N(R)\Bigr]\biggr\}\ ,
\label{5}
 \eeqn
 where $\vec R$ and $\vec r$ are the $\bar qq$ and $Gq$ transverse
separations respectively.  Making use of relation $\la r^2\ra \ll \la
R^2\ra$ one can keep only the second term in the curly brackets what makes
further calculations rather simple.
 
The cross section of gluon bremsstrahlung by a quark interacting with a
nucleon is found \cite{k3p1,k3p2} to be,
 \beq
\sigma^{qN}_{rad} = \sum\limits_{n=1}\frac{1}{n!}\,
\left[\frac{4\,\alpha_s}{3\,\pi}\,\,
{\rm ln}\left({s\over s_0}\right)\right]^n\,
\frac{9}{4}\,C\,r_0^2\ .
\label{6}
 \eeq
Since the mean separation $r_0$ is small, one can use the approximation
$\sigma_{\bar qq}(r)\approx C\,r^2$, where $C\approx 2.3$ can be evaluated
perturbatively. The QCD coupling $\alpha_s$ was averaged over the
radiation spectrum with a result $\alpha_s = 0.38 - 0.43$ in a good
agreement with the critical value $\alpha_c=0.43$ evaluated in
\cite{gribov}. 

To obtain the total $pp$ cross section one should add the cross section of
$pp$ interaction without any gluon radiation. This contribution,
$\sigma^{pp}_0$ is pure nonperturbative and cannot be evaluated reliably.
We treat it as a free parameter. Thus, the total $pp$ cross section takes
the form,
 \beq
\sigma^{pp}_{tot}= \tilde\sigma^{pp}_0 +
\frac{27}{4}\,C\,r_0^2\,\,
\left({s\over s_0}\right)^{\Delta}\ ,
\label{7}
 \eeq
with
 \beq
\Delta=\frac{4\,\alpha_s}{3\,\pi}
= 0.17 \pm 0.01\ ,
\label{8}
 \eeq
 and $\tilde\sigma^{pp}_0=\sigma^{pp}_0-9Cr_0^2/4$.

This value of $\Delta$ is about twice as large as the
one suggested by the data for the energy dependence
of total $pp$ cross 
sections when the simple parameterization $\sigma^{pp}_{tot}\propto
s^\Delta$ is applied. However, the
radiative part is a rather small fraction
of the total cross section (at medium high energies) and the presence
of the energy independent term $\sigma_0$ substantially reduces the
effective $\Delta_{eff}$.

\section{Unitarization and comparison with data}

The cross section (\ref{7}) apparently violates the Froissart bound and one
should introduce unitarity corrections \cite{dklt}.  Unfortunately, this is
not a well defined procedure since different recipes can be found in the
literature. 

The simplest known way to restore unitarity is to eikonalize the partial
amplitude $\gamma_{\pom}(b,s)$,
 \beq
{\rm Im}\,\Gamma_{\pom}(b,s)=
1 - {\rm exp}
\Bigl[-{\rm Im}\,\gamma_{\pom}(b,s)\Bigr]\ .
\label{6.16a}
 \eeq 
 At very high $s$ this amplitude approaches the black disk limit
\cite{dklt}, ${\rm Im}\,\Gamma_{\pom}(s,b) \to
\Theta\bigl[R^2(s)-b^2\bigr]$,
with radius, $R(s)=r_0\,\Delta\,{\rm ln}(s/s_0)$. Correspondingly, at
asymptotic energies,
 \beq
\Delta\,{\rm ln}\left({s\over s_0}\right) \gg
\frac{\la r_{ch}^2\ra}{r_0^2}\ ,
\label{6.17}
\eeq
 all hadronic cross sections reach the maximal universal energy growth 
allowed by Froissart-Martin's bound,
 \beq
\sigma^{hN}_{tot}(s) \to
2\,\pi\,r_0^2\,\Delta^2\,
{\rm ln}^2\left({s\over s_0}\right)\ .
\label{6.18}
\eeq
  
The eikonalization procedure (\ref{6.16a}) would be suitable if the
the incoming hadrons were eigenstates of the interaction \cite{zkl}.
Hadrons, however, are subject to diffractive off-diagonal excitation,
and the eikonal form of unitarization should be corrected in a way similar
to Gribov's inelastic corrections for hadron--nucleus
cross sections. The lowest order unitarity correction in (\ref{6.16a})
comes from the quadratic term in the exponent expansion of $\Gamma(b,s)$.
It has to be modified using the AGK cutting rules to include
single diffraction,
 \beqn
{\rm Im}\,\Gamma_{\pom} &=&
{\rm Im}\,\gamma_{\pom} -
{1\over2}\,\Bigl({\rm Im}\,\gamma_{\pom}\Bigr)^2\,
\Bigl[1\,+\,D(s)\Bigr]\nonumber\\
&+&\,  
O\bigl(\gamma_{\pom}^3\bigr)\ ,
\label{6.19}
\eeqn
 where $D(s) =\sigma_{sd}(s)/\sigma_{el}(s)$ is approximately $0.25$ in the
ISR energy range and decreases slightly with energy
$\propto s^{-0.04}$ \cite{dino} due to stronger unitarity corrections.
Asymptotically, as $s\to\infty$, $D(s)$ vanishes since
$\sigma_{el}(s) \propto {\rm ln}^2s$ and $\sigma_{sd}(s) \propto {\rm
ln}s$.

The inelastic corrections to higher order terms in the expansion
(\ref{6.16a}) are poorly known. A simple way to keep (\ref{6.19}) and to  
include diffraction into the higher terms is to modify (\ref{6.16a}) as,
 \beqn
&& {\rm Im}\,\Gamma_{\pom}(b,s)=
\frac{1}{1+D(s)}\nonumber\\
&\times& 
\left\{1 - {\rm exp}
\left[-\Bigl(1+D(s)\Bigr)\,
{\rm Im}\,\gamma_{\pom}(b,s)\right]
\right\}\ ,
\label{6.20}
\eeqn
which is known as quasi-eikonal model \cite{kaidalov}.

Following \cite{hp} we assume that the $t$-dependence of the $pp$ elastic
amplitude is given by the Dirac electromagnetic formfactor squared. 
Correspondingly, the mean square radius $\la \tilde r_{ch}^2\ra$ evaluated
in \cite{hp} is smaller than $\la r_{ch}^2\ra$. 

For the dipole parameterization
of the formfactor the partial elastic amplitude
which is related via unitarity to the $n$th term in
the radiation cross section (\ref{6})
takes the form,
\beq
{\rm Im}\,\gamma^{pp}_n(b,s)=
\frac{\sigma^{pp}_n(s)}{8\,\pi\,B_n}\,
y^3\,K_3(y)\ ,
\label{6.14}
\eeq
where $K_3(y)$ is the third order modified Bessel function
and $y=b\sqrt{8/B_n}$. The slope parameter grows
linearly with $n$ due to random walk of radiated gluons with a
step $r_0^2$ in the impact parameter plane,
$B_n = 2\la\tilde r_{ch}^2\ra/3 + n\,r_0^2$.

In order to calculate the total cross section, $\sigma_{tot}= 2\int
d^2b\,{\rm Im}\Gamma(b,s)$, one needs to fix the energy independent term
with $n=0$ in (\ref{7}).  This can be done comparing with the data for
$\sigma_{tot}$ at any energy sufficiently high to neglect Reggeons.
We use the most precise data \cite{cdf} at $\sqrt{s}=
546\,GeV$ and fix $\tilde\sigma_0=39.7\,mb$. 

The predicted energy dependence of $\sigma_{tot}^{pp}$ shown
by the dashed curve in Fig.~1
\begin{figure}[thb]
\includegraphics{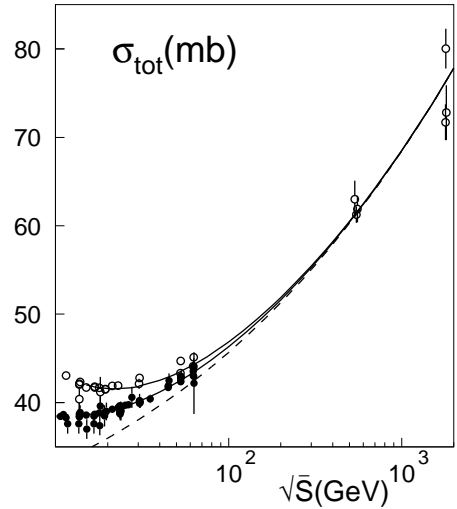}
\begin{center}
\vspace{5.5cm}
\parbox{7cm}
{\caption[shad1]
{Data for total $pp$ (full circles) and $\bar pp$ (open circles)
cross section 
and the prediction of Eq.~(\ref{6.14}) 
for the energy dependence of
the Pomeron part (dashed curve) and 
including Reggeon contribution (solid curves.}
\label{fig:1}}
\end{center}
\end{figure}
is in good
agreement with the data at high energies,
but apparently needs Reggeon corrections towards low energies.

In order to improve description of data at medium high energies (ISR) we
added the contribution of leading Reggeons parameterized in the standard
way with common parameters for $pp$ and $\bar pp$, the intercept fixed at
$\alpha_R(0)= 1/2$ and the slope parameter $\alpha_R^\prime=0.9\,GeV^{-2}$.
The Reggeons are added directly in the partial elastic amplitude,
 \beqn
{\rm Im}\,\Gamma(s,b) &=& {\rm Im}\,\Gamma_{\pom}(s,b) 
\nonumber\\ &+&
\Gamma_R(s,b)\,\Bigl[1\,-\,{\rm Im}\,\Gamma_{\pom}(s,b)\Bigr]\ ,
\label{6.15b}
\eeqn
with a proper screening by absorptive corrections.
The results are shown by the solid
curves of Fig.~1 ($pp$ bottom curve and $\bar pp$ upper curve). 

Employing Eq.~ (\ref{6.20}) we can also 
predict the slope of elastic scattering at $t=0$, $B_{el}(s) = \la
b^2\ra/2$, where averaging is weighted by the partial amplitude
(\ref{6.20}).  The results exhibit good agreement when compared with the
$pp$ and $\bar pp$ data in Fig.~2.
 \begin{figure}[thb]
\includegraphics{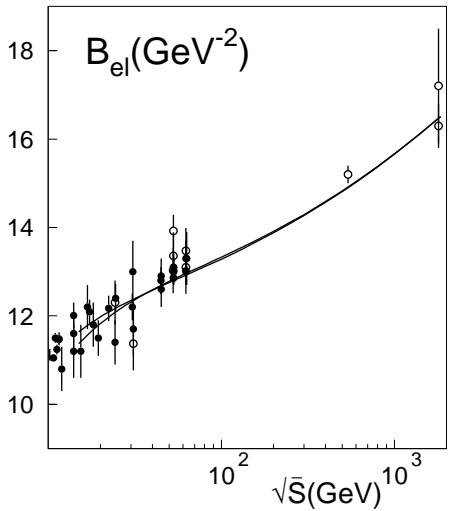}
\begin{center}
\vspace{5.5cm}
\parbox{7cm}
{\caption[shad1]
{Data for the elastic slope 
and our predictions. 
The upper and bottom curves and open and full circles 
correspond to $\bar pp$ and $pp$ respectively.}
\label{slope}}
\end{center}
\end{figure}

Note that the the slope essentially depends on
our choice of $\la\tilde r^2_{ch}\ra$ in (\ref{6.14}),
however the predicted energy dependence, {\it i.e.}
the effective value $\alpha_{\pom}^{\prime}$ is fully defined
by the parameter $r_0$ fixed in \cite{kst2}.
Indeed, each radiated gluon makes a small 
``step'' $\sim r_0^2 = 0.1\,fm^2$
in the impact parameter plane leading to the rising
energy dependence of the elastic slope.
Eventually, at very high energies the approximation
of small gluon clouds should break down. Nevertheless, the mean 
number of gluons in a quark 
$\la n\ra = \Delta\,{\rm ln}(s/s_0)$ remains quite small
in the energy range of colliders. 
It is only $\la n\ra = 0.7-1$ at the ISR 
and reaches about two gluons at the Tevatron.
Correspondingly, the mean square of the quark
radius grows from $0.06\,fm^2$ to  $0.18\,fm^2$ which
is still rather small compared to the mean square of 
the charge radius of the proton.

\section{Pomeron trajectory in impact parameter space}

Actually, we know more about the elastic partial amplitude than just
integral characteristics like the total cross section and slope,
we know the shape of $b$-dependence. To compare the $b$-distribution
and its development with energy directly with experimental data 
an analysis similar to one which has been done by Amaldi and Schubert
\cite{as} was performed recently in \cite{k3p2}.
The data for differential elastic cross section was fitted,
including both the real and imaginary parts of the amplitude, and then
Fourier transformed to impact parameter representation. As different from
\cite{as}, no model dependence was involved (geometrical scaling was
assumed in \cite{as}) and data from the ${\rm S\bar ppS}$ at
$\sqrt{s}=540\,GeV$ were included in addition to data from the ISR.
The data for partial elastic amplitude as function of impact parameter are
plotted in Fig.~3 in comparison with our predictions based on 
Eq.~(\ref{6.15b}).
 \begin{figure}[thb] 
\includegraphics{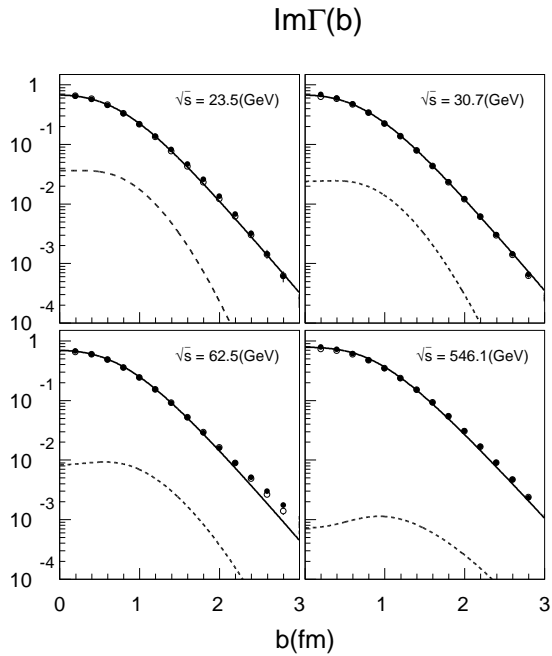}
\begin{center}
\vspace{7.5cm}  
\parbox{7cm}
{\caption[shad1]
{\sl The imaginary part of the partial
amplitude ${\rm Im}\,\Gamma(b)$ as function of
impact parameter at different c.m. energies. 
The solid curves are the theoretical prediction with Eq.~(\ref{6.15b}), 
and the dashed curves show the contribution of Reggeons.}
\label{gam-b}}
\end{center}
 \end{figure}
 Agreement between the data and our predictions is remarkably good,
especially if to recall that the Pomeron part has no free parameters,
except one, $\tilde\sigma_0$, adjusted to the total cross section measured
at one energy $\sqrt{s}=546\,GeV$ \cite{cdf}. Both the predicted shape of
the partial amplitude and its energy development are confirmed by the data. 

Addition of the $S\bar ppS$ data at $\sqrt{s}=540\,GeV$ into the analysis
plays a crucial role, it substantially increases the energy range important
for study of energy dependent effects. In particular it helps to extract
from the data the effective Pomeron trajectory $\alpha_{\pom}(b)=1 +
\Delta(b)$ as function of impact parameter, describing energy dependence
$s^{\Delta(b)}$ of the partial elastic amplitude.  The results are depicted
in Fig.~4.
 \begin{figure}[thb]
\includegraphics{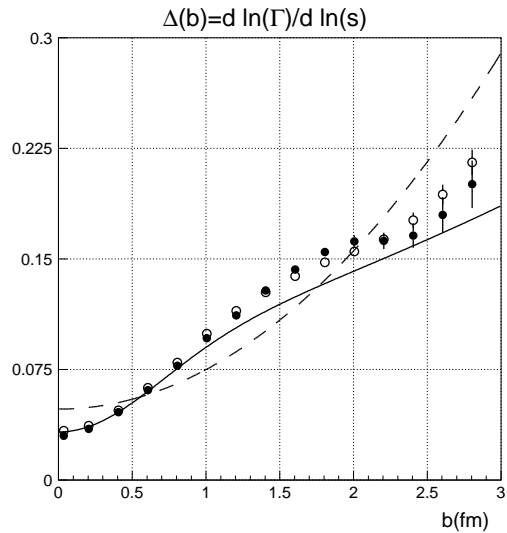}
\begin{center}
\vspace{6.5cm}
\parbox{7cm}
{\caption[shad1]
{\sl The effective Pomeron trajectory in impact parameter space.
Black and open points correspond to different parameterization
of the $t$-dependent elastic amplitude. 
Our prediction and of the single Pomeron pole model are shown by solid
and dashed curves respectively.}
\label{fig:2}}
\end{center}
 \end{figure} \vspace*{-0cm}
 The solid curves show the slope of energy dependence calculated with
Eq.~(\ref{6.15b}) without any adjustment. The agreement is rather good. 

The simple parameterization of the amplitude by a single Pomeron pole
without unitarity corrections fitted to data for total cross section and
elastic slope \cite{pdt} corresponds to the dashed curve.  Although this
parameterization is very successful describing total cross section it
poorly reproduces energy dependence of the partial amplitude at different
impact parameters. At the same time we should notice that the wide spread
prejudice that this model leads to a precocious breaking of unitarity at
small $b$ is incorrect. Wee see from Fig.~4 that $\Delta(b=0)$ predicted by
this simple model is rather small. In the case of geometrical scaling it
would be zero.

\section{Back to hard diffraction: Do we understand the results from
HERA?}

Diffraction of highly virtual photons, $\gamma^*\,p\to X\,p$ observed at
HERA is expected to be dominated by soft processes \cite{kp}.  Indeed, the
dominant {\it hard} fluctuations of the photon, have a small transverse
size $\sim 1/Q^2$, therefore the diffraction cross section is tiny, $\sim
1/Q^4$. On the other hand, {\it soft} fluctuations of large size which
appear very rarely with probability $\sim 1/Q^2$ (for transverse photons),
interact with a large hadronic cross section, therefore they dominate the
``hard'' DIS diffraction. Consequently, one might expect that
the cross section should rise with energy (at fixed $Q^2$)  like the
hadronic
cross sections, $\propto s^{0.08}$.  However, a much steeper slope of
energy dependence was observed at HERA.  According to the analysis in
\cite{royon} data from the H1 and ZEUS experiments lead to
 \beqn
\Delta &=& 0.2 \pm 0.02\ \ \ \ \ (H1)\ ,
\nonumber \\
\Delta &=& 0.13 \pm 0.04\ \ \ \ \ (ZEUS)\ .
\label{hera}
 \eeqn

The source of excitations with large effective masses is diffractive gluon
bremsstrahlung related via unitarity to the energy dependent second term in
the elastic amplitude (total cross section) in Eq.~(\ref{7}).  At the same
time, small mass diffraction is governed by excitation of the $\bar qq$
fluctuation in the photon without gluon radiation (yet it may
include radiation of energetic gluons). This diffractive component is
related by unitarity to the first energy independent term in Eq.~(\ref{7}).
Therefore, for large excitation masses we should expect about the same
value of $\Delta$ given by Eq.~(\ref{8}), rather than the averaged
effective value $\sim 0.08$ fitted to the total cross section.

Formulating differently, we can say that the softest part of interaction
responsible for the constant term in the total cross section Eq.~(\ref{7}) 
is related to the large size $R$ of the valence quark skeleton.  Such
interaction is too soft to resolve the gluon clouds of small size $r_0$. 
It needs a semi-hard interaction with transverse momentum $\sim 1\,GeV$ to
shake off the gluons from the quarks. This part of interaction is
responsible for the energy dependent term in (\ref{7}) and for diffractive
gluon radiation, {\it i.e.} excitation of large effective mass. 

\section{Summary}

Our main observation is existence of small gluonic clouds surrounding the
valence quarks in light hadrons. The most soft part of interaction with
transverse wave length longer than these spots cannot either resolve them or
shake them off the quarks. As an example, it can be treated in the string
model as a crossing and rearrangement of the hadronic strings.  The
corresponding part of the total cross section is energy independent since
the size of the valence quark skeleton does not grow with energy.

On the contrary, the gluon spots do rise with energy since the gluon
population increases and they perform a Braunian motion in the transverse
plane.  The gluon bremsstrahlung cross section also rises since each gluon
can be radiated at different rapidities with about the same probability. 
This is the source of energy dependence of the total cross section. 

Such a two-scale picture of interaction leads to presence of the two terms
in the total cross section Eq.~(\ref{7}). The first one is energy
independent, and the second one rises as power of energy with quite a large
exponent given by (\ref{8}). This second term is fully predicted, not only
its energy dependence, but also the normalization. Only the first
energy-independent term is treated as a free parameter and is fixed by
fitting just one experimental point. 

Our calculations reproduce well the energy dependence of the total
$pp$ cross section, and the elastic slope. Moreover, the data for
differential elastic scattering Fourier transformed to impact parameter
representation also well agree with our predictions, including the
effective Pomeron trajectory in impact parameter space. 

The two components of the total cross section can be well separated in
diffractive dissociation. Indeed, the small mass diffraction corresponds to
excitation of the quark ensemble without gluon radiation, {\it i.e.} it is
energy independent. Only diffractive excitation of gluon clouds leads to
the large mass dissociation, called in Regge phenomenology triple-Pomeron
diffraction.  This part is rather small, but steeply rises with energy with
the power given by (\ref{7}). This indeed has been observed by experiments
at HERA.

We have found a new regime where one can perform calculations employing the
smallness of the size $r_0=0.3\,fm$ of the gluonic spots compared to large
interquark separation $R\sim 1\,fm$. On the contrary, in DIS one deals with
an opposite limit when the gluon clouds are much larger than the interquark
separation. In that case the cross section
can be estimated using perturbative QCD provided that 
$Q^2\gg Q_0^2$ ($Q_0\sim 1\,GeV$), and $s\gg s_0$ ($s_0\sim
1\,GeV^2$), but $x=Q^2/s\ll 1$. Dependent on the approximation used, two
models for the Pomeron are known, the BFKL \cite{bfkl} and
double-leading-log approximation. The total virtual
photoabsorption cross section is predicted to rise steeply with energy as
is confirmed by data from HERA.

Our conclusion about smallness of gluonic spots in light hadrons is in a
good accord with what has been already observed in models for
nonperturbative QCD vacuum.  It has been found employing the QCD sum rules
\cite{braun} that the gluonic formfactor of the proton has very weak $Q^2$
dependence corresponding to a small radius of the gluon distribution
$\sim 0.3\,fm$. A small gluon correlation radius $\sim 0.3\,fm$
also appears in the lattice calculations \cite{pisa}. It is also predicted
by the model of instanton liquid \cite{shuryak}. Experimental observation
of a small cross section for large mass soft diffractive dissociation which
has led the analyses \cite{kst2} to a small value of $r_0$ can be treated
more generally, in particular as a confirmation for a small size cloud of
any kind of gluonic vacuum fluctuations surrounding the valence quarks.
They are usually referred to as constituent quarks. 

Note that the two-scale structure of the energy-dependent total cross
section has to exist also in other models attempting to incorporate the
nonperturbative effects \cite{kl,kkl}.

\medskip

\noi
{\bf Acknowledgment:} I am grateful to the Organizing Committee
for kind invitation and financial support.

\end{document}